\documentclass[12pt]{article}

\usepackage{amsmath,amsfonts,amssymb}

\usepackage{epsf}

\newcommand{\be}{\begin{equation}}
\newcommand{\ee}{\end{equation}}
\newcommand{\beq}{\begin{eqnarray}}
\newcommand{\eeq}{\end{eqnarray}}
\newcommand{\bea}[2]{\be\label{#2}\begin{array}{#1}}
\newcommand{\eea}{\end{array}\ee}

\def\lfig#1#2#3#4{
 \begin{figure}
 \refstepcounter{figure}
 \label{#4}
 \addtocounter{figure}{-1}
 \epsfxsize=#3
 \centerline{\epsfbox{#2}}
 {\bf \caption{{\rm #1}}}
 \end{figure}
}

\def\Rb{{\rm \bf R}}

\def\Im{\,{\rm Im}\, }
\def\Re{\,{\rm Re}\, }

\def\rangl{\right\rangle   }
\def\langl{\left\langle  }
\def\({\left(}
\def\){\right)}
\def\[{\left[}
\def\]{\right]}
\def\p{\partial}
\def\D{\slash\!\!\!\! D}
\def\11{1\!\! 1}

\def\hf{{1\over 2}}

\def\e{\epsilon}
\def\eps{\varepsilon}

\def\vph{\varphi}
\def\L{\Lambda}
\def\S{\Sigma}

   \def\CA {{\cal A}}
   
   \def\CC {{\cal C}}

   \def\CK {{\cal K}}
   \def\CL {{\cal L}}
   \def\CM {{\cal M}}
   
   \def\CO {{\cal O}}

   \def\CS {{\cal S}}
   \def\CT {{\cal T}}

   \def\CX {{\cal X}}
   
   \def\CZ {{\cal Z}}

\def\bX{\bar X}
\def\bF{\bar F}

\def\bZ{\bar Z}

\def\bb{\bar b}

\def\bw{\bar w}

\def\bz{\bar z}

\def\bv{\bar v}

\def\bzt{\bar \zeta}

\def\gst{g_{\rm s}}

\def\vr{{\vec r}}

\def\flat{0}

\def\etaf{\eta^{\flat}}
\def\vf{v^{\flat}}
\def\bvf{\bv^{\flat}}
\def\xf{x^{\flat}}
\def\rf{r^{\flat}}
\def\vrf{\vr^{\flat}}
\def\wf{w_{\flat}}
\def\bwf{\bw_{\flat}}
\def\wcf{w_{\rm c}}
\def\bwcf{\bar w_{\rm c}}
\def\psif{\psi_{\flat}}

\def\zf{z^{\flat}}
\def\bzf{\bz^{\flat}}

\def\cf{c_{\flat}}

\def\em{\epsilon^-}
\def\ep{\epsilon^+}

\def\et{\epsilon^3}

\def\etap{\eta_+}
\def\etam{\eta_-}
\def\etapm{\eta_\pm}

\def\ztp{\zeta_+}
\def\ztm{\zeta_-}
\def\ztpm{\zeta_\pm}

\def\bztp{\bzt_+}
\def\bztm{\bzt_-}

\def\bztmp{\bzt_\mp}

\def\Schi{\chi}
\def\KTw{K_{\CZ}}

\def\MHM{\CM}
\def\MHKC{\CS}

\def\dil{\Rb^{\times}}

\begin{document}

\title{Quantum covariant c-map}

\author{Sergei Alexandrov\thanks{email: Sergey.Alexandrov@lpta.univ-montp2.fr}
}

\date{}

\maketitle

\vspace{-0.8cm}

\begin{center}
\it{Laboratoire de Physique Th\'eorique \& Astroparticules} \\
\it{Universit\'e Montpellier II, 34095 Montpellier Cedex 05, France}

\end{center}

\vspace{0.3cm}

\begin{abstract}
We generalize the covariant c-map found in hep-th/0701214 including perturbative
quantum corrections.
We also perform explicitly the superconformal quotient from the hyperk\"ahler cone obtained by the
quantum c-map to the quaternion-K\"ahler space, which is the moduli space of hypermultiplets.
As a result, the perturbatively corrected metric on the moduli space
is found in a simplified form comparing to the expression known in the literature.
\end{abstract}

\section{Introduction}

Compactifications of type II superstring theories on a Calabi-Yau manifold lead to
low energy effective theories which consist of two independent sectors
represented by vector multiplets (VM) and hypermultiplets (HM) coupled
to $N=2$ supergravity.
The two sectors are decoupled from each other in accordance with factorization
of the moduli space of the Calabi-Yau to the complex structure and the K\"ahler structure moduli.
Despite of this decoupling, at the tree level
the corresponding effective actions appearing in the string compactifications can be
related by the so called c-map \cite{Cecotti:1988qn,Ferrara:1989ik}. It allows to construct
a non-linear $\sigma$-model for the hypermultiplets in type IIA (IIB) theory
from the holomorphic prepotential, completely characterizing the vector multiplets,
of type IIB (IIA) theory.
Its origin can be traced back to the T-duality in a 3-dimensional theory
obtained by a compactification on a circle of the original 4-dimensional
effective theory.

Recently an important progress has been achieved in understanding and generalizing the c-map.
First of all, it was formulated off-shell in terms of $N=2$ projective superspace \cite{Rocek:2005ij}
(see also earlier works \cite{Berkovits:1995cb,Berkovits:1998jh}
where superspace effective actions in relation
with the c-map were analyzed).
The main advantage of this formulation is that it describes the effective action for HM
in terms of a single function.
More precisely, the general target space for the HM $\sigma$-model is a quaternion-K\"ahler space $\MHM$
of real dimension $4n$ where $n-1=h_{1,1} (h_{1,2})$ in type IIA (IIB) theory \cite{Bagger:1983tt}.
Over such space one can always construct the so called Swann bundle $\MHKC$,
known also as hyperk\"ahler cone, which is a K\"ahler space of dimension $4(n+1)$ possessing a quaternion structure
and a homothetic Killing vector. Physically it represents the target space of a superconformal extension
of the original HM $\sigma$-model. In the case when $\MHM$ has $n+1$ commuting isometries,
which is always true at the perturbative level,
both $\sigma$-models can be dualized to a theory of tensor multiplets (TM). The latter has a very elegant
description in terms of a single holomorphic function $G$ on $N=2$ superspace, known as ``generalized prepotential".
Thus, all the complicated geometry of the quaternion-K\"ahler manifold turns out to be encoded in this function.
The work \cite{Rocek:2005ij} provided a simple relation between the generalized prepotential $G$
and the holomorphic prepotential $F$ for the vector multiplets.
Besides, it opened an avenue to explore profound connections between the hypermultiplet geometry
and the black hole physics \cite{Pioline:2006ni}.

But this is only the beginning of the story because, contrary to the VM sector,
the HM sector receives both perturbative and non-perturbative corrections
in the string coupling constant $\gst$. Whereas only partial results exist about non-perturbative
contributions to the HM geometry (see for example
\cite{Becker:1995kb,Davidse:2005ef,Alexandrov:2006hx,Robles-Llana:2006is}), the perturbative corrections
were completely understood in \cite{Robles-Llana:2006ez} (see also
\cite{Antoniadis:1997eg,Antoniadis:2003sw,Anguelova:2004sj}). The projective superspace description
mentioned above played a crucial role in this construction and the result was nicely formulated
in terms of a simple correction to the classical generalized prepotential $G$. This allowed to talk
about ``quantum c-map".

There is however one drawback inherent to both treatments \cite{Rocek:2005ij} and \cite{Robles-Llana:2006ez}.
All calculations in these two works were done in a gauge, which fixes the superconformal symmetry.
For the purposes of \cite{Rocek:2005ij,Robles-Llana:2006ez}, which were
to derive the quaternion-K\"aler metric starting from the generalized prepotential,
this gauge fixing was sufficient
and in fact it simplified a lot this procedure known as superconformal quotient.
However, some geometric aspects of the construction, which may and do have some physical applications,
remained hidden. In particular, the gauge fixing complicates the search for
non-perturbative corrections to the hyperk\"ahler cone and the generalized prepotential.

At the tree level this drawback was overcome in the recent work \cite{Neitzke:2007ke} where the so called
covariant c-map was constructed and applied to the radial quantization of BPS black holes.
Here we are going to generalize this result in a natural way by constructing a ``{\it quantum}
covariant c-map". This means that we perform the superconformal quotient starting from the generalized
prepotential found in \cite{Robles-Llana:2006ez}
without a gauge fixing and explicitly find coordinates on $\MHM$ as functions on $\MHKC$ invariant under
the dilatations and $SU(2)_R$ transformations.

This study benefits us in several ways. First, we improve the formulae for the covariant c-map
from \cite{Neitzke:2007ke} not only including the quantum corrections, but also making them regular
in the limit where the gauge used in the previous works is imposed. Second, we reveal an anomaly
in conformal transformations of some quantities due to the quantum correction. Its possibility was
missed in the general treatment of the tensor/hypermutliplet duality \cite{deWit:2001dj} and
here we fulfill this gap.
Finally, the metric on the moduli space of the hypermultiplets, which we obtain after the superconformal quotient,
is much simpler than the one found in \cite{Robles-Llana:2006ez}.
The reason is that we perform the superconformal quotient directly for the hypermultiplets,
whereas in \cite{Robles-Llana:2006ez} it was done at the level of the tensor multiplets and only after that
the resulting action was dualized to the HM $\sigma$-model.
Although the two results are equivalent, their form is quite different.

The organization of the paper is as follows. First, we briefly review the derivation of a quaternion-K\"aler
metric with $n+1$ commuting isometries from a generalized prepotential.
In section \ref{sec_covSCQ} we apply this procedure to the particular case where the generalized
prepotential is given by the quantum c-map \cite{Robles-Llana:2006ez}.
All calculations here are done avoiding any gauge fixing and the results culminate in the formulae
for the quantum covariant c-map. Then in section \ref{sec_hyperm} we reproduce the perturbatively
corrected metric on the moduli space. For this purpose it is sufficient to work in a gauge, which we use to simplify
the derivation. Section \ref{sec_sum} is devoted to a summary of the main results.

\section{Quaternion-K\"ahler geometry from projective superspace}
\label{sec_review}

In this section we review the relation between quaternion-K\"ahler spaces
and the projective superspace formalism \cite{Karlhede:1984vr,Gates:1984nk,Hitchin:1986ea,Lindstrom:1987ks}.

The projective superspace is a convenient way to write (off shell) $N=2$ conformally invariant supersymmetric actions.
In the given case we are interested in the action for $N=2$ tensor multiplets.
It can be written as the following integral\footnote{With few exceptions we follow the notations and
the normalizations of \cite{Rocek:2005ij},}
\be
S_{\rm TM}=\int d^4x\, d^2\theta\, d^2\bar\theta \,\CL,
\qquad
\CL(v,\bv,x)=\Im \oint_{\CC}{d\zeta\over 2\pi i\zeta}\, G\(\eta(\zeta),\zeta\),
\label{contint}
\ee
where
\be
\eta^I=\frac{v^I}{\zeta}+x^I-\bv^I\zeta
\label{o2mult}
\ee
are ``real $\CO(2)$ projective superfields", which are written in terms of $N=1$ chiral superfields $v^I$
and $N=1$ real linear superfields $x^I$. The index $I$ enumerates the multiplets
and in our case runs from 0 to $n$. The function $G$ is called ``generalized prepotential" and, together
with the contour $\CC$, completely determines the model. The constraints from superconformal invariance
restrict $G$ to be a function homogeneous of first degree in $\eta^I$ and without explicit dependence of
$\zeta$ \cite{deWit:2001dj}. In turn, this induces a set of constraints on the superspace
Lagrangian density $\CL$. In particular, it must also be homogeneous of first degree.

After eliminating the auxiliary fields, one remains with two scalars, which we denote as the corresponding
$N=1$ superfields they come from, and a tensor gauge field $B_{\mu\nu}$ with the field strength
$H^\mu=-\hf \eps^{\mu\nu\rho\sigma}\p_{\nu}B_{\rho\sigma}$. The action for these fields is
\be
S_{\rm TM} = \int d^4 x\[ \CL_{x^I x^J}\(\p_\mu v^I\p^\mu \bv^J
+\frac{1}{4}\(\p_\mu x^I\p^\mu x^J-H_\mu^I H^{\mu J} \)\)
+\frac{i}{2}\(\CL_{v^I x^J}\p_\mu v^I-\CL_{\bv^I x^J}\p_\mu \bv^I \)H^{\mu J}\].
\ee
In 4 dimensions the antisymmetric field can be dualized to a scalar. This is achieved by
adding to the action a term $\frac{i}{2}(w_I-\bw_I)\p_\mu H^{\mu I}$.
The real part of $w_I$ is determined by
\be
w_I+\bw_I=\p_{x^I}\CL.
\label{wxCL}
\ee
Thus, $H_{\mu}^I$ can be eliminated by means of equations of motion, whereas $x^I$ can be
found as functions of $w_I$, $v^I$ and their conjugates through \eqref{wxCL}.
As a result, this leaves a $\sigma$-model for
$(v^I,\bv^I,w_I,\bw_I)$ which form $n+1$ hypermultiplets. The constraints on $\CL$ ensure
that the target space for this $\sigma$-model is a hyperk\"ahler cone $\MHKC$ with
the hyperk\"ahler potential given by the Legendre transform of $\CL$ \cite{deWit:2001dj}
\be
\Schi(v,\bv,w,\bw)=\CL(v,\bv,x)-(w_I+\bw_I)x^I.
\label{legtr}
\ee
The quaternion structure on $\MHKC$ is formed by three complex structures. The first
one is canonical, {\it i.e.}, ${J^{3p}}_q=i\delta^p_q$,
where the indices $p,q$ run over both $v^I$ and $w_I$, and the other two
are given by ${J^{+\bar p}}_q=g^{\bar p r}\Omega_{rq}$ and its complex conjugate where
the holomorphic two-form is
\be
\Omega=dw_I\wedge dv^I.
\label{Omtwo}
\ee

The hyperk\"ahler cone can be reduced to a quaternion-K\"aler space by means of
superconformal quotient. A useful concept in this procedure, which
plays an intermediate role, is a twistor space $\CZ$.
It has one complex dimension less and is a K\"ahler quotient of $\MHKC$.
To construct $\CZ$, it is enough to notice that $\chi$ is homogeneous of first degree
in $v^I$ and $\bv^I$ and is invariant under their $U(1)$ rotations.
Therefore, one can single out one coordinate, say $v^n$, and define
complex coordinates on $\CZ$ as $(w_I,z^A)$ where
\be
z^I=(z^A,1)=\frac{v^{I}}{v^n},\qquad A=0,\dots, n-1.
\ee
Then the K\"ahler potential $\KTw$ on the twistor space is determined from
the factorization of $\chi$
\be
\Schi=\sqrt{v^n\bv^n}\,e^{\KTw}.
\label{chiKT}
\ee

This twistor space is an $S^2$ fibration over $\MHM$ and, to perform the remaining quotient,
it is convenient to fix a gauge. We will impose the gauge used in
\cite{Rocek:2005ij,Robles-Llana:2006ez}, which is $\vf=0$ (which becomes $z^0=0$ on $\CZ$).
Then the metric on the underlying quaternion-K\"ahler space is given by
\be
g_{\alpha\bar\beta}=
-4\(\p_\alpha \p_{\bar \beta}\KTw- e^{-2\KTw}\CX_{\alpha}\bar \CX_{\bar\beta}\),
\label{metsuper}
\ee
where $\alpha,\beta$ label holomorphic coordinates on $\MHM$, $(w_I,z^a)$, $a=1,\dots n-1$,
and $\CX$ is a holomorphic one-form coming out from the holomorphic two-form $\Omega$ \eqref{Omtwo}.
In the chosen gauge it reads as \cite{Rocek:2005ij}
\be
\CX_{\zf=0}=2d w_n+2z^a d w_a.
\label{Xone}
\ee
Thus, we conclude that starting from the generalized prepotential $G$, evaluating the contour integral
\eqref{contint}, doing the Legendre transform \eqref{legtr} and performing the superconformal quotient,
one arrives at the quaternion-K\"ahler metric for the target space of the HM $\sigma$-model.

Finally, let us notice that since the hyperk\"ahler potential $\Schi$ depends on $w$ only through
the combination $w+\bw$, it is evident that $\MHKC$ possesses $n+1$ commuting triholomorphic isometries.
These isometries descend to $\MHM$ and preserve there the quaternion
structure \cite{deWit:2001dj}.

\section{Quantum c-map, hyperk\"aler cone and twistor space}
\label{sec_covSCQ}

\subsection{The perturbed prepotential}

As was shown in \cite{Robles-Llana:2006ez}, the quantum c-map can be nicely summarized
by the following relation between the holomorphic prepotential $F$, determining the special
K\"ahler geometry of the VM sector, and the generalized prepotential
\be
G(\eta)={F(\eta^\L)\over \etaf}+4ic\etaf\ln \etaf.
\label{qcmap}
\ee

Just to summarize the necessary information, we recall that
$F$ is a homogeneous of second degree function of $n$ variables labeled by $\L\in \{1,\dots n\}$.
On a rigid special K\"ahler manifold it defines the K\"ahler potential and the metric as
\be
K=i\(\bX^\L F_\L(X)-X^\L \bar F_\L(\bX)\),
\qquad
N_{\L\S}=i\(F_{\L\S}(X)-\bar F_{\L\S}(\bX)\),
\ee
where $X^{\L}$ are homogeneous coordinates and $F_\L,F_{\L\S}$ denote derivatives
of the prepotential $F$ with respect to $X^{\L}$. $N^{\L\S}$ will denote
the inverse of the metric $N_{\L\S}$.
The corresponding quantities for the local special
geometry are given by
\be
\CK=\ln\(Z^\L N_{\L\S}\bZ^\S\),
\qquad
\CK_{a\bar b}=\p_a \p_{\bar b}\CK,
\ee
where derivatives are already evaluated with respect to projective coordinates $Z^a=X^a/X^n$.

The first term in \eqref{qcmap} describes the classical c-map and was determined in \cite{Rocek:2005ij}.
The second term gives the one-loop correction. The constant $c$ is given by the Euler number of
the Calabi-Yau
\be
c=-\frac{\chi}{12}=-\frac{1}{6\pi}\(h_{1,1}-h_{1,2}\).
\ee
Here we wrote this term for the type IIA theory. In the type IIB case it is enough to change the sign
of $c$. The authors of \cite{Robles-Llana:2006ez} also argued that there are no higher loop corrections.
Thus, the generalized prepotential \eqref{qcmap} is our starting point to get the hyperk\"ahler cone
and the quaternion-K\"ahler space for the hypermultiplets at the perturbative level.

\subsection{Choice of the contour and the tensor Lagrangian}
\label{subsec_cont}

The first thing one needs to do is to evaluate the superspace Lagrangian density $\CL$ given by
the integral \eqref{contint}. But before doing this, one should choose a contour of integration.
With the function $G(\eta)$ given in \eqref{qcmap}, the integrand has the following singularities:

i) poles at $\zeta=0,\ztp,\ztm$ where $\ztp$ and $\ztm$ are roots of $\zeta \etaf$ given by
\be
\ztpm=\frac{\xf\mp\rf}{2\bvf}, \qquad
\rf=\sqrt{(\xf)^2+4\vf\bvf}\,;
\ee

ii) logarithmic singularities at $\zeta=0,\ztp,\ztm,\infty$ which must be joined by two cuts.

\lfig{The singularities, logarithmic cuts and contour $\CC$ on the complex $\zeta$-plane.}
{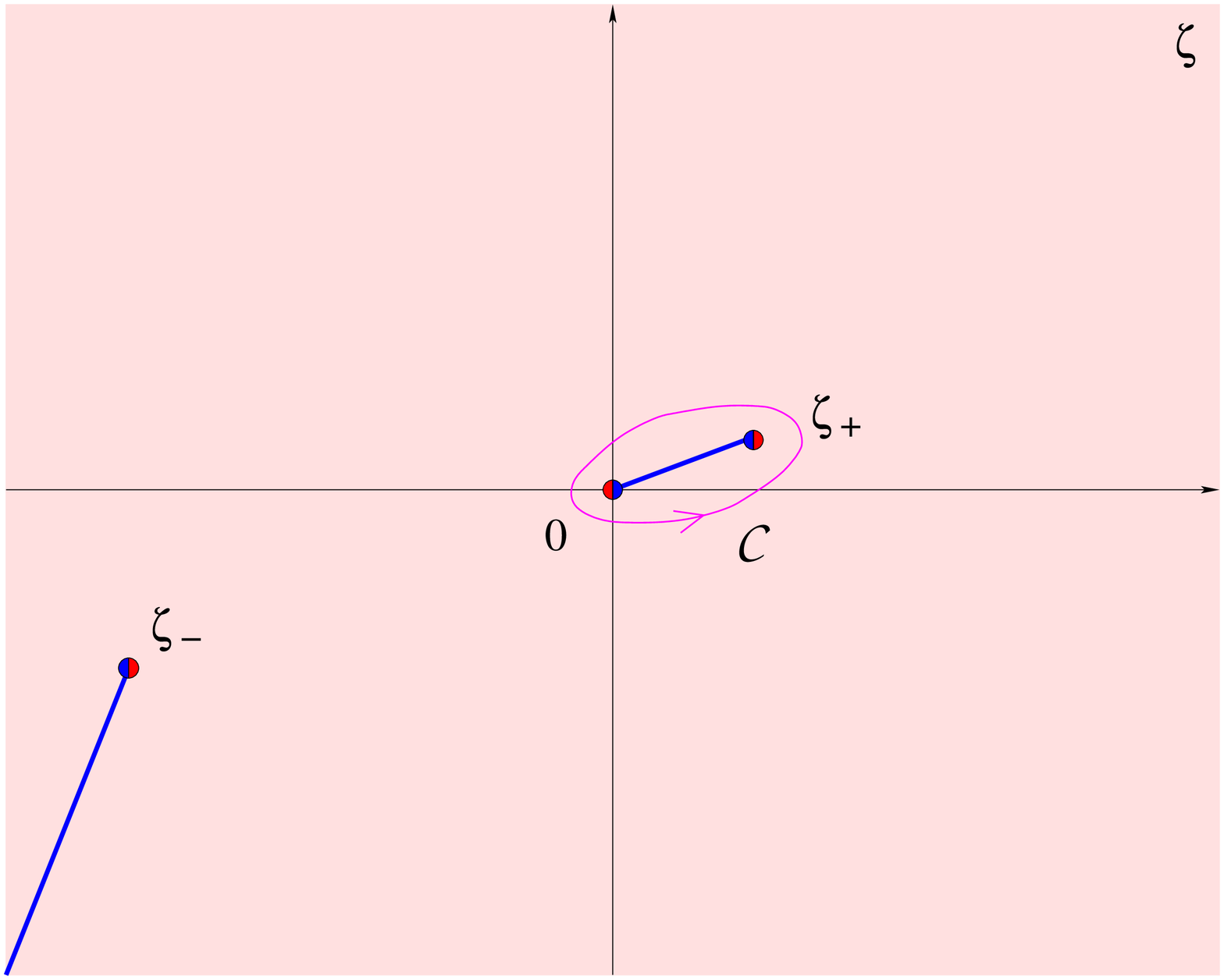}{9cm}{contour}

\noindent
From \cite{Rocek:2005ij,Robles-Llana:2006ez} we know also that in the gauge $\vf=0$, where
$\ztp\to 0$ and $\ztm\to\infty$, the contour encircling the origin gives the correct result.
Therefore, it is natural to demand that our contour $\CC$ reduces to such circle in this limit.
This leaves the only possibility, which is to take the two logarithmic cuts along $(0,\ztp)$
and $(\ztm,\infty)$, respectively, and to choose $\CC$ around the first cut as depicted
in fig. \ref{contour}.\footnote{In fact, it is possible to choose different contours for
different terms of the generalized prepotential. This possibility is indeed realized in
various applications (see for example \cite{Ivanov:1995cy,Houghton:1999hr}).
In particular, the so called ``figure-eight" contour \cite{Hitchin:1986ea},
which encircles $\ztp$ and $\ztm$ in the opposite directions,
was shown to be appropriate for the quantum correction term \cite{Anguelova:2004sj}.
However, we have two reasons to not follow this possibility.
First, the ``figure-eight" contour would generate additional terms diverging in the
limit $\vf\to 0$ (see below). Secondly, we expect that the full non-perturbative
generalized prepotential can be summed up to a function possessing some special properties.
In particular, in the type IIB case there should a trace of the modular invariance
of the hyperk\"ahler potential constructed in \cite{Robles-Llana:2006is}. Since the modular
transformations affect the string coupling, from our point of view it is unnatural to separate terms
of different degree in $g_s$.
}

The evaluation of the contour integral presented in appendix \ref{ap_integral}
leads to the following result for the tensor
Lagrangian
\be
\CL(v,\bv,x)=\frac{1}{\rf}\,\Im F(\etap)-\xf\Im\frac{F(v)}{(\vf)^2}+x^\L\Im\frac{F_\L(v)}{\vf}
+4c\(\xf-\rf+\xf\ln\frac{\xf+\rf}{2}\),
\label{tenslagr}
\ee
where we denoted
\beq
\etapm^\L\equiv\eta^\L(\ztpm)&=&
x^\L-v^\L\bztmp-\bv^\L\ztpm
\nonumber \\
&=& x^\L-\frac{\xf}{2}\(\frac{v^\L}{\vf}+\frac{\bv^\L}{\bvf}\)
\pm\frac{\rf}{2}\(-\frac{v^\L}{\vf}+\frac{\bv^\L}{\bvf}\).
\eeq

Let us notice that the contribution of the pole at $\zeta=0$ from the first term in \eqref{qcmap}
is linear in $x^I$. Therefore, it will not contribute to the Legendre transform \eqref{legtr}
and this is the reason why the authors of \cite{Neitzke:2007ke} have chosen the contour
which encircles only $\ztp$. However, it does contribute to the definition of variables on $\MHKC$ and $\MHM$
and, if one misses this contribution, various quantities
become singular in the limit $\vf\to 0$ so that this gauge is not achievable anymore.
On the other hand, the presence of the quantum correction and, as a consequence, of the logarithmic cut
requires for the contour to go also around the origin and cancels all singularities.

\subsection{Legendre transform and K\"ahler potential}

To perform the Legendre transform and to find the hyperk\"ahler potential, we follow the strategy
applied in \cite{Neitzke:2007ke}. First, let us do the Legendre transform in $x^\L$ keeping $\xf$
untouched. This gives
\beq
\langl \CL-(w_\L+\bw_\L)x^\L\rangl_{x^\L} &=&
\frac{K(\etap,\etam)}{4\rf}-\frac{\xf K(v,\bv)}{4\vf\bvf}
-\frac{\xf}{2}\(\frac{v^\L}{\vf}+\frac{\bv^\L}{\bvf}\)(w_\L+\bw_\L)
\nonumber \\
&&
+4c\(\xf-\rf+\xf\ln\frac{\xf+\rf}{2}\),
\eeq
where the relation between $x^\L$ and $w_\L+\bw_\L$ reads
\be
w_\L+\bw_\L\equiv \p_{x^\L}\CL=\Im\(\frac{1}{\rf}\,F_\L(\etap)+\frac{1}{\vf}\,F_\L(v)\).
\ee

To proceed further, it is convenient to define
the variables playing the role of electric and magnetic potentials
\beq
\vph^\L &\equiv& \frac{i}{2}\(\frac{v^\L}{\vf}-\frac{\bv^\L}{\bvf}\)
=\frac{1}{\rf}\Im \etap^\L,
\\
\psi_\L &\equiv& w_\L+\bw_\L-\Im\frac{F_\L(v)}{\vf}=\frac{1}{\rf}\Im F_\L(\etap).
\label{tildchi}
\eeq
These relations are similar to the attractor equations determining the asymptotic moduli
in terms of the charges for BPS black holes in $N=2$ supergravity
\cite{Ferrara:1995ih,Ferrara:1996um,Ooguri:2004zv}.
In the same way we can use them to express $\etap^\L$ in terms of $\vph^\L$ and $\psi_\L$
as
\be
\etap^\L=\rf\(i\vph^\L-\frac{\p\Sigma(\vph,\psi)}{\p\psi_\L}\),
\qquad
F_\L(\etap)=\rf\(i\psi_\L+\frac{\p\Sigma(\vph,\psi)}{\p\vph^\L}\),
\ee
where
\be
\Sigma(\vph,\psi)=\frac{K(\etap,\etam)}{4(\rf)^2}
\ee
is the so called Hesse potential of rigid special K\"ahler geometry \cite{Hitchin:1999qf,LopesCardoso:2006bg}.

Now it is easy to take the Legendre transform in $\xf$ because both $\vph^\L$ and $\psi_\L$
are independent of it. Thus, one finds
\be
\Schi\equiv \langl \CL-(w_\L+\bw_\L)x^\L-(\wf+\bwf)\xf\rangl_{x^\L,\xf}
=\frac{4\vf\bvf}{\rf}\,\Sigma(\vph,\psi)-4c\rf,
\label{HKpot}
\ee
where $\xf$ is determined as a function of other variables by
\be
\frac{\xf}{\rf}\,\Sigma(\vph,\psi)+4c\ln\frac{\xf+\rf}{2}=\psif
\label{xchif}
\ee
with
\be
\psif = \wf+\bwf+\hf\(\frac{v^\L}{\vf}+\frac{\bv^\L}{\bvf}\)(w_\L+\bw_\L)+\frac{K(v,\bv)}{4\vf\bvf}-4c.
\label{tildchi0}
\ee
This gives the loop corrected hyperk\"ahler potential. The quantum correction
appears in two places. First, it comes in a simple form as a term linear in $\rf$. Secondly, it
changes the function $\rf(v,\bv,w+\bw)$ itself due to the additional term in \eqref{xchif}.
This term makes the equation for $\xf$ irrational and, contrary to the classical case,
it cannot be solved explicitly.

The hyperk\"ahler potential is invariant under the Peccei-Quinn symmetries generated by the following
Killing vectors
\be
K=-\frac{i}{2}\(\p_{\wf}-\p_{\bwf}\),
\quad
P^\L=\frac{i}{2}\(\p_{w_\L}-\p_{\bw_\L}\),
\quad
Q_\L=-\vf\p_{v^\L}+w_\L\p_{\wf}-\frac{F_{\L\S}(v)}{2i}\,\p_{w_\S}+{\rm c.c.}
\label{isom}
\ee
These isometries are triholomorphic and therefore descend to quaternionic isometries
on $\MHM$. As it is clear from \eqref{isom}, they are not affected by the quantum correction.
However, the generators $Q_\L$ are affected by the inclusion of the terms
insuring finiteness in the $\vf\to 0$ limit.

The passage to the twistor space goes in a simple way. As we discussed in section \ref{sec_review},
it is enough to single out, for example, $v^n$ and project it out. Then
the K\"ahler potential on the twistor space can be determined from \eqref{chiKT}
and is given by
\be
\KTw=\ln \[\frac{\zf\bzf}{\rho}\,
\Sigma\(\frac{i}{2}\(\frac{z^\L}{\zf}-\frac{\bz^\L}{\bzf}\),w_\L+\bw_\L-\Im\frac{F_\L(z)}{\zf}\)
-c\rho\]+\ln 2,
\label{KpotCZ}
\ee
where $\rho$ is defined through the following equation
\beq
& \sqrt{1-\frac{4\zf\bzf}{\rho^2}}\,
\Sigma\(\frac{i}{2}\(\frac{z^\L}{\zf}-\frac{\bz^\L}{\bzf}\),w_\L+\bw_\L-\Im\frac{F_\L(z)}{\zf}\)
+4c\ln\frac{\sqrt{\rho^2-4\zf\bzf}+\rho}{2}
&
\nonumber \\
& =\wcf+\bwcf +\hf\(\frac{z^\L}{\zf}+\frac{\bz^\L}{\bzf}\)\(w_\L+\bw_\L\)+\frac{K(z,\bz)}{4\zf\bzf}-4c  &
\eeq
and we had to change $\wf$ by
\be
\wcf=\wf-2c\ln v^n.
\label{wcw}
\ee
The underling reason for the latter change will become clear in the next subsection where
we discuss the transformation properties under the superconformal group.

From the above expressions one can see that projecting out $\vf$ instead of $v^n$, one could obtain
more simple formulae. We have chosen $v^n$ to have possibility to consider the limit $\vf\to 0$
also on the twistor space. This gauge will be used to evaluate the metric on the HM moduli space
in section \ref{sec_hyperm}.

\subsection{Conformal transformations and anomaly}

Before discussion of the covariant c-map, the subject of the next subsection,
one has to establish transformation properties of all variables under the dilations and
the $SU(2)_R$ symmetry, which are fixed or projected out
by the superconformal quotient.

For the tensor multiplets these transformations are known be
\bea{cccc}{su2xv}
\dil: &\ &
v^I\to \mu^2 v^I, &
x^I\to \mu^2 x^I,
\\
\rule{0pt}{12pt}
SU(2): &\ &
\delta v^I=i\et v^I+\ep x^I,
\quad
\delta \bv^I=-i\et v^I+\em x^I,
& \quad
\delta x^I=-2\(\ep \bv^I+\em v^I\)
\eea
so that $\vr^I=\(x^I,2v^I,2\bv^I\)$ have scaling weight 2 and transform as a three vector.
Then the general analysis \cite{deWit:2001dj} leads to conclusion that on the hyperk\"ahler cone
$w_I$ are invariant under dilations and have the
the following transformations under $SU(2)$
\be
\delta w_I=\ep \CL_{v^I}, \qquad
\delta \bw_I=\em \CL_{\bv^I}.
\label{gentrw}
\ee
However, it is easy to see that these rules are inconsistent with the equation \eqref{xchif},
relating $\xf$ and $\wf$, and with the $su(2)$ algebra.

The reason for the failure of the general results presented above can be traced back to
the failure of the homogeneity of the generalized prepotential $G(\eta)$ \eqref{qcmap}
due to the presence of the logarithmic correction.
The same quantum correction destroys the homogeneity of the tensor Lagrangian \eqref{tenslagr}
and suggests to weaker the homogeneity condition in the following way
\be
x^I \CL_{x^I}+v^I \CL_{v^I}+\bv^I \CL_{\bv^I}=  \CL+c_I x^I,
\label{hompropan}
\ee
where $c_I$ are some real constants. In our case
\be
c_I=4c\,\delta_{I}^{\flat}.
\label{CIour}
\ee
The condition \eqref{hompropan} was indeed found in \cite{deWit:2001dj}
(see appendix A there) as the most general one
insuring the conformal invariance of the TM action. However, it was argued that
the terms in $\CL$ giving rise to $c_I$ can be neglected since they do not
contribute neither to the TM action, nor to the hyperk\"ahler potential on $\MHKC$.
Nevertheless, it turns out that these terms affect the relations between tensor and hypermultiplet
variables and the conformal transformations of the latter. Therefore, we reconsider here
the derivation of these transformations.

The transformation law under dilatations for $w_I$ can be found from their definition,
$w_I+\bw_I=\CL_{x^I}$. The condition \eqref{hompropan} together with \eqref{su2xv}
implies that
\be
w_I \to w_I+c_I \ln \mu.
\label{dilw}
\ee
The SU(2) transformations are determined from the requirement of invariance of
the action obtained by dualization of the antisymmetric field $B_{\mu\nu}$
from the tensor multiplets to a scalar. As we mentioned in section \ref{sec_review},
the dualization amounts the addition of the term
$\frac{i}{2}(w_I-\bw_I)\p_\mu H^{\mu I}$.
The transformations \eqref{gentrw}
allow to cancel all the terms appearing after the $SU(2)$ transformations
from the original TM action, which arise since $\p_\mu H^{\mu I}$ vanishes now only on-shell.
But this still leaves a freedom to add
to $\delta w_I$ a constant imaginary term, which will contribute only a total derivative.
The precise constant can be fixed by the $su(2)$ algebra.
One gets
\be
\delta^{(3)} w_I=-\frac{1}{2i}\[\delta^{(+)},\delta^{(-)}\]w_I
=\frac{1}{2i}\,\delta^{(-)}\CL_{v^I} =-\frac{c_I}{2i}.
\ee
Thus, the full $SU(2)$ transformation on the hyperk\"ahler cone obtained
from the tensor Lagrangian satisfying \eqref{hompropan} reads
\be
\delta w^I=\ep \CL_{v^I}-\frac{c_I}{2i}\,\e^3, \qquad
\delta \bw^I=\em \CL_{\bv^I}+\frac{c_I}{2i}\,\e^3.
\label{trwbw}
\ee

Now we can understand the nature of the change \eqref{wcw}. The logarithmic term
cancels the anomalous terms in the transformations of $\wf$ appearing
due to non-vanishing $\cf=4c$ \eqref{CIour}. The new variable $\wcf$ is invariant under
both dilatations and $U(1)$ transformations associated with the $\e^3$-generator.
Thus, it is a natural variable on the twistor space which is obtained from the
hyperk\"ahler cone by a quotient along these two symmetries.
Its $SU(2)$ transformations are given by
\be
\delta \wcf=\ep \(\CL_{\vf}-2c\,\frac{x^n}{v^n}\), \qquad
\delta \bwcf=\em \(\CL_{\bvf}-2c\,\frac{x^n}{\bv^n}\).
\ee
This shows that only the $\ep$-generator acts non-trivially on $\wcf$, as it would be the case for $w_I$
when there are no anomalous terms in the homogeneity condition.

One can show that the anomaly in the homogeneity condition
does not change the complex structures on $\MHKC$.
In particular, the holomorphic two-form $\Omega$ is still given by \eqref{Omtwo}.
Upon reduction to the twistor space, it gives rise to the following holomorphic forms
\be
\omega=dw^c_A\wedge dz^A,
\qquad
\CX=2\(dw^c_n+z^A dw^c_A\)-c_A dz^A,
\label{holforms}
\ee
where we introduced coordinates generalizing \eqref{wcw},
$w^c_I=w^c_I-\hf\,c_I \ln v^n$. In the gauge $\vf=0$ the one-form $\CX$
reduces to
\be
\CX_{\zf=0}=2d w_n^c+2z^a d w_a^c-c_a dz^a.
\label{Xonemod}
\ee
However, since in our particular case all $c_\L$ vanish, the last term disappears and
$w^c_\L$ appearing in \eqref{Xonemod} coincide with the usual $w_\L$.
As a result, the one-form contributing to the perturbative metric on the HM moduli space
does not differ from the standard expression \eqref{Xone}.

\subsection{The covariant c-map}
\label{sec_covcmap}

Here we give explicit results for the quantum covariant c-map, {\it i.e.},
functions on the hyperk\"ahler cone $\MHKC$, which are invariant with respect to
the transformations discussed in the previous subsection and play the role of coordinates
on the quaternion-K\"ahler space obtained by the quantum c-map.
Essentially, we need just a small generalization of the corresponding expressions
found in \cite{Neitzke:2007ke} to include the quantum correction
and to take into account the terms coming from the contribution of the pole $\zeta=0$
to the classical tensor Lagrangian (see discussion in section \ref{subsec_cont}).
Besides, one should remember about the anomalous terms in \eqref{trwbw}.
As a result, one arrives at the following invariant functions:
\beq
& \displaystyle
e^{\phi}\equiv \frac{\Schi}{4\rf}=\frac{\vf\bvf}{(\rf)^2}\,\Sigma(\vph,\psi)-c,
& \nonumber \\
& \displaystyle
Z^a \equiv \frac{\etap^a}{\etap^n}, \qquad a=1,\dots,n-1,
& \nonumber \\
& \displaystyle
A^\L \equiv \frac{1}{(\rf)^2}\(\vrf\cdot \vr^\L\)
=\hf\(\frac{v^\L}{\vf}+\frac{\bv^\L}{\bvf}\)+\frac{\xf}{(\rf)^2}\Re\etap^\L
, &
\label{coor}
\\
& \displaystyle
B_\L\equiv
-i(w_\L-\bw_\L)+\frac{\xf}{(\rf)^2}\, \Re F_\L(\etap)+\Re\frac{F_\L(v)}{\vf},
& \nonumber \\
& \displaystyle
\sigma \equiv
i(\wf-\bwf)+\frac{i}{2}\(\frac{v^\L w_\L}{\vf}-\frac{\bv^\L \bw_\L}{\bvf}\)
-\frac{\xf}{2(\rf)^2}\, \Re\(\etap^\L B_\L-A^\L F_\L(\etap) \)-2ic\ln\frac{\etap^n\vf}{\etam^n\bvf}.
\nonumber
\eeq
Their invariance can be checked using explicit SU(2) transformations of various quantities
presented in Appendix \ref{ap_su2tr}.

The Peccei-Quinn generators \eqref{isom} acting on the space spanned by these functions have precisely
the same form as in the classical case \cite{Neitzke:2007ke}
\be
K=\p_{\sigma},
\quad
P^\L=\p_{B_\L}-\hf\,A^\L\p_{\sigma},
\quad
Q_\L=-\p_{A^\L}-\hf\,B_\L\p_{\sigma}.
\label{isomQK}
\ee
They are true isometries of the quaternion-K\"ahler metric which
will be derived in the next section.

\section{Loop corrected HM moduli space}
\label{sec_hyperm}

In this section our aim is to derive the metric on the HM moduli space
including the perturbative quantum corrections.
This was already done in \cite{Robles-Llana:2006ez}, but here we apply a different strategy.
Instead of performing the superconformal quotient at the level of tensor Lagrangians and then
dualizing to hypermultiplets, we first pass to the hyperk\"ahler cone and perform the quotient
following the procedure described in section \ref{sec_review}.
The resulting quaternion-K\"ahler metric will be equivalent to the one found in
\cite{Robles-Llana:2006ez}, but with some essential simplifications.

In fact, the first part of the program has been already completed in the previous section.
We can start directly from the K\"ahler potential \eqref{KpotCZ} on the twistor space.
Then, according to the procedure of section \ref{sec_review}, it remains to impose the gauge $\vf=0$
and to evaluate the metric \eqref{metsuper}. The result should be expressed in terms of
the coordinates given by the covariant c-map. Since we need only their expressions in the fixed gauge,
first we discuss the limit $\vf\to 0$ in some detail.

\subsection{The $\vf\to 0$ limit}

It is trivial to compute all quantities in this limit taking into account the
following expansion
\be
\etap^\L\approx -\frac{\xf}{\vf}\, v^\L+x^\L+\frac{\vf}{\xf}\,\bv^\L-\frac{\bvf}{\xf}\,v^\L
\ee
and the homogeneity property of the holomorphic prepotential.
In particular, one finds:
\beq
\mbox{the tensor Lagrangian}
& \quad &
\CL={1\over 4x^0}\(N_{IJ}(v,\bv)x^Ix^J -2K(v,\bv)\)
-4c x^0\ln x^0,
\\
\mbox{the hyperk\"ahler potential on $\MHKC$}
& \quad &
\Schi \approx \frac{K(v,\bv)}{\xf(v,\bv,w,\bw)}-4c\xf(v,\bv,w,\bw),
\\
\mbox{the K\"ahler potential on $\CZ$}
& \quad &
\KTw=
\ln \( {K(z,\bz)\over \rho(z,\bz,w,\bw) } -4c\rho(z,\bz,w,\bw) \),
\eeq
where $\xf(v,\bv,w,\bw)$ is defined through
\be
{K(v,\bv)\over 2(\xf)^2}-4c(\ln \xf+1)
=(w_\L+\bw_\L)N^{\L\S}(w_\S+\bw_\S)-(\wf+\bwf),
\ee
whereas $\rho(z,\bz,w,\bw)$ is given by a similar equation with the replacements
$v$ by $z$ and $\wf$ by $\wcf$. These functions are simply related as $\rho=\xf/\sqrt{v^n\bv^n}$.
The invariant coordinates \eqref{coor}
reduce in this limit to the following expressions
\beq
&
\displaystyle
e^{\phi}\approx \frac{K(v,\bv)}{4(\xf)^2}-c,
\qquad
Z^a\approx
\frac{v^a}{v^n}=
z^a,
\qquad
A^\L\approx \frac{x^\L}{\xf},
&
\nonumber \\
&
\displaystyle
B_\L\approx
\vphantom{-i(w_\L-\bw_\L)+\frac{x^\S}{\xf}\Re F_{\L\S}(v)=}
-i(w_\L-\bw_\L)+A^\S\Re F_{\L\S}(v),
&
\label{limcoor}
\\
&
\displaystyle
\sigma \approx
\vphantom{2i(\wf-\bwf)+i(w_\L-\bw_\L)\frac{x^\L}{\xf}- 4ic\ln\frac{v^n}{\bv^n}}
i(\wcf-\bwcf)-\hf\,A^\L B_\L +\hf\, A^\L A^\S \Re F_{\L\S}(v).
\nonumber
&
\eeq
For $c=0$ they coincide with the coordinates introduced in \cite{Neitzke:2007ke}.\footnote{The only
difference is an overall factor 2 in $A^\L$ and $\sigma$.} Notice also that in terms of these
coordinates the K\"ahler potential $\KTw$ has an explicit and very simple form
\be
\KTw=\hf\(\CK(Z,\bZ)+\ln\frac{4\,e^{2\phi}}{e^{\phi}+c}\).
\label{KTreal}
\ee

We emphasize that one obtains well defined expressions for all quantities in this limit.
This is opposite to the situation in \cite{Neitzke:2007ke}, where, for example, $\CL$ and
$B_\L$ diverge. The regular behavior is achieved by including the contribution of the
pole $\zeta=0$ to the classical part of the tensor Lagrangian. If one does not do this, in the
limit $\vf\to 0$ the contour in \eqref{contint}
is pinched between the two poles, $\zeta=0$ and $\zeta=\ztp$,
which is the reason for divergences.
The inclusion of the quantum correction automatically requires for the contour
to encircle both poles and makes everything regular.

\subsection{The quternion-K\"ahler metric}

Now it is straightforward to evaluate the metric on the quaternion-K\"ahler space.
It is given by \eqref{metsuper} with the one-form $\CX$ from \eqref{Xone}.
It is more convenient however to express it in terms of the coordinates \eqref{limcoor}.
Since the change of coordinates is not holomorphic, it is a bit tedious calculation.
As an intermediate step, we present the inverse of the transformation \eqref{limcoor} and
the derivatives of the twistor potential $\KTw$
with respect to the original holomorphic coordinates on $\MHM$ in Appendix \ref{ap_metric}.
As a result, one obtains the following metric on the HM moduli space
\beq
ds^2&=&{r+2c \over r^2(r+c)}\,d r^2
-{1\over r}\(N^{\L\S}-{2(r+c)\over r K}\,Z^\L\bZ^\S\)
\(F_{\L\Theta}dA^\Theta-dB_\L\)\(\bF_{\S\Xi}dA^\Xi-dB_\S\)
\nonumber \\
&&+ {r+c \over 4 r^2(r+2c)}\(d\sigma-\hf \(A^\L dB_\L-B_\L d A^\L\)-2c\, d\CA\)^2
-{4(r+c)\over r}\,\CK_{a{\bar b}}\,dZ^a d\bZ^{\bar b}.
\label{hypmet}
\eeq
where we introduced
\be
r=e^{\phi},\qquad
d\CA=i\(\CK_a dZ^a -\CK_{\bar a}d\bZ^{\bar a}\).
\ee

The form of the result \eqref{hypmet} is much simpler than the one found in \cite{Robles-Llana:2006ez}.
Nevertheless one can show that they are equivalent.
The key ingredient of the proof is the inverse of the matrix $\CT^{\rm qc}_{\L\S}$ introduced in
\cite{Robles-Llana:2006ez} (eq. (4.14)). In that paper it was not found due to
a complicated form of the original matrix, whereas here it can be read off directly from the metric
\eqref{hypmet} and is given by
\be
\(\CT^{\rm qc}\)^{\L\S}=\frac{4}{r}\(\frac{r+c}{rK}\(Z^\L\bZ^\S+Z^\S\bZ^\L\)-N^{\L\S}\).
\ee

\section{Summary}
\label{sec_sum}

The main results of this paper are

i) the {\it quantum} covariant c-map \eqref{coor},

ii) the {\it simplified} loop corrected metric on the hypermultiplet moduli space \eqref{hypmet}.

\noindent
Besides, we found an anomaly in conformal transformations of the coordinates on the
hyperk\"ahler cone defined by the Legendre transform. The anomaly is related to the failure
of the homogeneity due to the quantum correction. The modified dilatations and
SU(2) rotations are given in \eqref{dilw} and \eqref{trwbw}.

These results can be considered first of all as a groundwork to include non-perturbative
corrections. In particular, for the case of the universal hypermultiplet the non-perturbative
corrections are completely known in the one-instanton approximation \cite{Alexandrov:2006hx}.
Our results might be useful to formulate these corrections at the level of the
generalized prepotential $G(\eta)$, where however the simple $\CO(2)$ multiplets \eqref{o2mult}
are not enough anymore and more general $N=2$ multiplets must be taken into consideration
(see discussion in \cite{Anguelova:2004sj}). Once the corresponding function $G$ is found,
it may be used to generate all higher orders of the instanton expansion.

Another potential application of this work is related to BPS black holes.
Although the vector multiplets, which are usually used to describe supersymmetric black holes,
do not receive string loop corrections, it is interesting whether our results have some interpretation
in the black hole physics. Notice, in particular, that
at the tree level both hyperk\"ahler potential and K\"ahler potential on the twistor space
can be explicitly expressed in terms of the Hesse potential \cite{Neitzke:2007ke}, which
is known to provide the black hole entropy. In our case such explicit expressions do not exist,
but the relation \eqref{HKpot} for $\Schi$ looks simple enough to appeal for
an interpretation. It hints that $c\, \frac{(\rf)^2}{\vf\bvf}$ might be considered
as a correction to the entropy. However, its meaning remains absolutely unclear.

\section*{Acknowledgements}
It is a pleasure to thank Stefan Vandoren for very valuable discussions.
The research of the author is supported by CNRS and by the contract
ANR-05-BLAN-0029-01.

\appendix

\section{Evaluation of the tensor Lagrangian}
\label{ap_integral}

Our aim is to evaluate
\be
\CL(v,\bv,x)=\Im \oint_{\CC}{d\zeta\over 2\pi i\zeta}\({F(\eta^\L)\over \etaf}+4ic\etaf\ln \etaf\)
\ee
with the contour $\CC$ shown in fig. \ref{contour}.
To disentangle the simple pole and the logarithmic singularity at $\zeta=0$
in the second term, it is convenient to shift one of them by a small $\eps$
and to take the limit $\eps\to 0$ after the evaluation of the integral.
Thus, we have
\beq
\CL(v,\bv,x)&=&\mathop{\rm lim}\limits_{\eps\to 0}\Im\oint_{\CC}{d\zeta\over 2\pi i}
\[-\frac{F(\zeta\eta^\L)}{\bvf\zeta^2(\zeta-\ztp)(\zeta-\ztm)}
\right.
\nonumber \\
&&\qquad \qquad \left.
-\frac{4ic\bvf}{\zeta^2}(\zeta-\ztp)(\zeta-\ztm)\ln\(\bvf(\zeta-\ztm)\(\frac{\ztp}{\zeta-\eps}-1\)\)\].
\nonumber
\eeq
The first, classical term gets contributions from the residues at $\zeta=0$ and $\ztp$.
Together they result in
\be
\CL_{\rm cl}(v,\bv,x)=\frac{1}{\rf}\,\Im F(\etap)-\xf\Im\frac{F(v)}{(\vf)^2}+x^\L\Im\frac{F_\L(v)}{\vf}.
\label{Lcl}
\ee
The second, quantum term picks up also two contributions: from the pole at $\zeta=0$ and
from the logarithmic cut along $(\eps,\ztp+\eps)$. They can be written as
\beq
\CL_{\rm q}(v,\bv,x)&=&4c \,\mathop{\rm lim}\limits_{\eps\to 0}\Re\[
\left. -\bvf\frac{\p}{\p\zeta}\((\zeta-\ztp)(\zeta-\ztm)\ln\(\bvf(\zeta-\ztm)\(\frac{\ztp}{\zeta-\eps}-1\)\)\)\right|_{\zeta=0}
\right.
\nonumber \\
&& \left.
+\bvf\int_{\eps}^{\ztp+\eps} d\zeta\, \frac{(\zeta-\ztp)(\zeta-\ztm)}{\zeta^2}\].
\nonumber
\eeq
Elementary calculations give
\be
\CL_{\rm q}(v,\bv,x)=4c\Re\[2\bvf\ztp+\bvf(\ztp+\ztm)\ln(\bvf\ztm)\].
\label{Lq}
\ee
Altogether \eqref{Lcl} and \eqref{Lq} result in the tensor Lagrangian \eqref{tenslagr}.

\section{$SU(2)$ transformations}
\label{ap_su2tr}

Here we list some of the SU(2) transformations
useful to check the invariance of the coordinates \eqref{coor}.
In particular, one has
\be
\delta \ztpm=-\ep -\em\ztpm^2+i\e^3\ztpm ,
\qquad
\delta \etapm^\L=-\(\ep \bztmp+\em\ztpm\)\etapm^\L.
\nonumber
\ee
Due to this the Hesse potential transforms homogeneously
\be
\delta\Sigma(\vph,\psi)=-\(\frac{\ep}{\vf}+\frac{\em}{\bvf}\)\xf\,\Sigma(\vph,\psi).
\nonumber
\ee
For $\psi_\L$ and $\psif$
the transformations can be derived directly from \eqref{tildchi} and \eqref{xchif}
and are given by
\beq
\delta \psi_\L &=&-\frac{1}{2i\rf}\[\ep\(\bztm F_\L(\etap)-\bztp\bF_\L(\etam)\)
+\em\(\ztp F_\L(\etap)-\ztm\bF_\L(\etam)\)\],
\nonumber
\\
\delta\psif &=&-\[\ep\(\frac{(\xf)^2}{\vf}+2\bvf\)+\em\(\frac{(\xf)^2}{\bvf}+2\vf\)\]
\frac{\Sigma(\vph,\psi)}{\rf}
-8c\, \frac{\ep\bvf+\em\vf}{\xf+\rf}.
\nonumber
\eeq
One can check that they are consistent with the transformations
of $w_\L$ and $\wf$ following from the generalized law \eqref{trwbw}
\beq
\delta w_\L &=&\ep\[-\frac{1}{2i\rf}\(\bztm F_\L(\etap)-\bztp\bF_\L(\etam)\)
+\frac{x^\S}{2i\vf}\,F_{\L\S}(v)-\frac{\xf}{2i(\vf)^2}\,F_\L(v)\],
\nonumber
\\
\delta \wf &=&
\ep\[  -\frac{\bvf}{i(\rf)^3}\(F(\etap)-\bF(\etam)\)
\right.
\nonumber \\
&+&
\frac{1}{2i(\rf)^2}\( \(\frac{(\xf)^2+\xf\rf+2\vf\bvf}{2(\vf)^2}\,v^\L+\bv^\L\)F_\L(\etap)
+\(\frac{(\xf)^2-\xf\rf+2\vf\bvf}{2(\bvf)^2}\,\bv^\L+v^\L\) \bF_\L(\etam)\)
\nonumber \\
&+& \left. \frac{\xf}{i(\vf)^3}\,F(v)
-\frac{x^\L}{2i(\vf)^2}\,F_\L(v)
-8c \,\frac{\bvf}{\xf+\rf}
\]+2ic\e^3.
\nonumber
\eeq

\section{Derivatives of $\KTw$}
\label{ap_metric}

The relations inverse to the change of coordinates \eqref{limcoor} can be summarized as follows
(recall that $r=e^{\phi}$)
\beq
& w_\L = {1\over 2i}\(F_{\L\S} A^\S- B_\L\), &
\nonumber
\\
& \wcf=
-r-c\(\ln {4(r+c)\over K}-1\)+\frac{1}{2i}\(\sigma+\hf A^\L B_\L-\hf\,A^\L A^\S F_{\L\S}\). &
\nonumber
\label{changew}
\eeq
Using these relations and the K\"ahler potential \eqref{KTreal}, one can obtain
the following derivatives of $\KTw$, which is considered here
as a function of $Z^a,w_\L,\wcf$ and their conjugates:
\beq
\p_{\wcf}\p_{\bwcf}\KTw&=& -\frac{r+c}{8r^2(r+2c)},
\nonumber \\
\p_{w_\L}\p_{\bwcf}\KTw &=&\frac{r+c}{8r^2(r+2c)}\, A^\L ,
\nonumber \\
\p_{Z^a}\p_{\bwcf}\KTw &=& \frac{r+c}{32 r^2(r+2c)}\(8c\CK_a -i\p_a F_{\L\S}A^\L A^\S\),
\nonumber \\
\p_{w_\L}\p_{\bw_\S}\KTw &=&\frac{1}{2r}\, N^{\L\S}-\frac{r+c}{8r^2(r+2c)}\, A^\L A^\S ,
\nonumber \\
\p_{Z^a}\p_{\bw_\L}\KTw &=&-\frac{i}{4r}\, N^{\L\S}\p_a F_{\S\Theta} A^{\Theta}
-\frac{r+c}{32 r^2(r+2c)}\, A^\L \( 8c\CK_a -i\p_a F_{\S\Theta}A^\S A^\Theta\),
\nonumber \\
\p_{Z^a}\p_{\bZ^{\bb}}\KTw &=& \frac{r+c}{2r}\, \CK_{a\bb}
-\frac{c^2}{2r^2}\, \frac{r+c}{r+2c}\, \CK_a\CK_{\bb}
+\frac{1}{8r}N^{\Theta\Xi}\p_a F_{\L\Theta}\p_{\bb}\bF_{\S\Xi} A^\L A^\S
\nonumber \\
&&+\frac{r+c}{16 r^2(r+2c)}\, A^\L A^\S \( ic\(\CK_{\bb} \p_a F_{\L\S}-\CK_{a} \p_{\bb} \bF_{\L\S}\)
-\frac{1}{8}\, \p_a F_{\L\Theta} \p_{\bb} \bF_{\S\Xi} A^{\Theta} A^{\Xi}\).
\nonumber
\eeq

\end{document}